\documentstyle[aas2pp4]{article}
\input epsf
\def\plotone#1{\centering \leavevmode
\epsfxsize= 0.8\columnwidth \epsfbox{#1}}

\def\etal{{\frenchspacing\it et al.}}

\def\rn{\noindent\parshape 2 0truecm 8.8truecm 0.3truecm 8.5truecm}
\def\nn#1 #2{#1, #2.}				
\def\nnn#1 #2 #3{#1, #2. #3.}			
\def\nnnn#1 #2 #3 #4{#1, #2. #3. #4.}		
\def\dualand{, \&\hbox{ }}				
\def\multiand{, \&\hbox{ }}				
\def\rfprep#1;#2;#3 {{\par\rn#1 #2, preprint #3\par}}

\def\rg#1;#2;#3;#4;#5;#6 {\par\rn#1 #2, {\it #3}, {\bf #4}, #5 (``#6'') \par}
\def\rf#1;#2;#3;#4;#5 {\par\rn#1 #2, {\it #3}, {\bf #4}, #5\par}
\def\rfbook#1;#2;#3;#4;#5 {{\frenchspacing\par\rn#1 #2, {\it #3} (#4: #5)\par}}
\def\rfproc#1;#2;#3;#4;#5;#6 {{\frenchspacing\par\rn#1 #2, in {\it #3}, ed. #4 (#5: #6)\par}}

\def\be{\begin{equation}}
\def\ee{\end{equation}}
\def\bea{\begin{eqnarray}}
\def\eea{\end{eqnarray}}

\def\muk{\mu{\rm K}}

\def\cmm2{{\,\rm cm^{-2}}}
\def\cm2{{\,{\rm cm}^2}}
\def\cmm3{{\,{\rm cm}^{-3}}}
\def\gcmm3{{\,{\rm g\,cm^{-3}}}}

\def\la{\mathrel{\mathpalette\fun <}}

\def\fun#1#2{\lower3.6pt\vbox{\baselineskip0pt\lineskip.9pt
  \ialign{$\mathsurround=0pt#1\hfil##\hfil$\crcr#2\crcr\sim\crcr}}}
\def\p{^\prime}
\def\bul{\noindent $\bullet$}

\begin{document}
\twocolumn
\title{Forecasting Foreground Impact on Cosmic Microwave Background 
Measurements}

\author{Lloyd Knox\footnote{e-mail:  {\it knox@flight.uchicago.edu}}}
\affil{University of Chicago}
\authoraddr{Department of Astronomy and Astrophysics, Chicago, IL 60637, USA}

\vspace{.2in}

\begin{abstract}
We explore the possible impact of galactic and extragalactic
foregrounds on measurements of the cosmic microwave background (CMB).
We find that, given our present understanding of the foregrounds, they
are unlikely to qualitatively affect the ability of the MAP and Planck
satellites to determine the angular power spectrum of the CMB, the key
statistic for constraining cosmological parameters.  Sufficiently far
from the galactic plane, the only foregrounds that will affect power
spectrum determination with any significance are the extragalactic
ones.  For MAP we find the most troublesome foregrounds are radio
point sources and the thermal Sunyaev-Zeldovich (SZ) effect.  For
Planck they are these same radio point sources and the Far Infrared
Background.  Prior knowledge of the statistics of the SZ
component (either via theoretical calculation, or higher frequency
observations of just a few percent of the sky, such as will be done by
balloon-borne experiments) may significantly improve MAP's
determination of the CMB power spectrum.  We also explore the
foreground impact on MAP and Planck polarization power spectrum
measurements.
\end{abstract}


\section{Introduction}

Much attention has been paid in recent years to the nature of the
foregrounds which obscure our view of the background.  This attention
has resulted in discoveries about the nature of the foregrounds as
well as methods for estimating CMB anisotropy from
foreground-contaminated data.  From studying these developments, we
have concluded that for planned large area, multi-frequency
experiments, such as MAP\footnote{MAP home page:
{\tt http://map.gsfc.nasa.gov}}
and Planck\footnote{Planck home page:\\ 
{\tt http://astro.estec.esa.nl/SA-general/\linebreak[0]Projects/Cobras/cobras.html}}, the foregrounds are unlikely to
be responsible for qualitative degradation of the primary cosmological
results.

This happy situation is due to a number of factors.  First, there is a
window in frequency space, where, at high galactic latitude, CMB
fluctuations are the brightest diffuse source in the sky.  Second, the
high-latitude galactic foregrounds are very smooth; they do not have
much small-scale fluctuation power.  Third, foregrounds, unlike
instrument noise and some systematic error sources, are suppressed at
small angular scales by the finite
angular resolution of the telescope.  Fourth, point source count
estimates suggest that only a small fraction of pixels in the MAP and
Planck maps will be affected---and these can be identified with
threshold cuts and removed.  Finally, even if uncertainty in a
particular mode of the CMB map is dramatically increased by the
presence of foregrounds, uncertainty in the CMB power spectrum may not
be significantly affected.  This is due to the fact that, at least in
the foreground free case, the dominant source of power spectrum
uncertainty (except at the smallest angular scales) comes from sample
variance, not instrument noise.

The primary cosmological results---determination of cosmological
parameters---depend mostly on how well the power spectrum is measured.
We thus focus on the impact of foregrounds on the determination of
this power spectrum.  Our method for estimating this impact can be
considered to be a generalization of those based on Wiener filtering
by Bouchet, Gispert, and Puget (1995, hereafter ``BGP95'') and Tegmark
and Efstathiou (1996, hereafter ``TE96'') as well as that of White
(1998, hereafter ``W98'').  All these approaches take the CMB and
foregrounds to be statistically isotropic, Gaussian-distributed
fields.  Given this assumption, estimation of the power spectrum
errors is straightforward, as described below.

As is always the case with parameter estimation, how well the desired
parameters can be recontsructed depends on the assumed prior
information.  The methods of TE96 and W98 essentially assume that the
foreground power spectra are known with infinite precision {\it a
priori}.  The most important difference between our method and theirs
is that we only assume finite prior information about the foreground
power spectra.

Although the method of BGP95 was derived assuming Gaussianity, they
have tested it with non-Guassian simulations of Planck Surveyor maps
(see, e.g., Gispert and Bouchet, 1996).  Their results lend
credibility to the forecasts derived analytically under the Gaussian
assumption.

Below we first describe our methods for estimating the power spectrum
uncertainties given an experimental configuration and foreground
model.  In section III we describe our model of the foregrounds, which
is based on that detailed in the Planck Phase A proposal (Bersanelli
\etal ~1996), and the High Frequency Instrument (HFI) and Low Frequency
Instrument (LFI) proposals\footnote{These proposals are not
yet publically available. Most of the work referred to here will soon
be available as Bouchet and Gispert (1998).}.

To date, foregrounds have been essentially ignored in estimates
of the cosmological parameter uncertainties\footnote{A notable exception
is the ``LDB'' case in Bond, Efstathiou \& Tegmark (1997) which was based on calculations by
the TopHat group of CMB pixel errors after pixel-by-pixel subtraction
of foregrounds in their Antarctic maps.}.  We find that
they are unlikely to qualitatively change the results.  Although
for MAP this conclusion depends somewhat on the amplitude of
the contribution from the Sunyaev-Zeldovich effect, which is not yet
sufficiently well-determined.

Not only does the amplitude of the SZ power spectrum affect the 
ability of MAP data to constrain the CMB power spectrum, but so does our
prior knowledge of it.  This is fortunate, because while the amplitude is
completely out of our control, we {\it can} do something about
how well we know it.  We emphasize that the prior information we
need is not of the actual SZ map, but of the statistics of the map.  
The statistics can be calculated theoretically, or by actually
measureing the SZ map over only a few per cent of the sky.

Higher order moments of the probability distribution may also be of
interest if the CMB statistics are non-Gaussian, which they will be to
some degree even if the primordial fluctuations are Gaussian.  Therefore,
we also estimate how well the amplitudes of individual
spherical harmonics can be determined.  The uncertainty on these
amplitudes is much more strongly affected by the presence of 
foregrounds than is the uncertainty on the power spectrum.

Effects due to contributions not in ones model of the data may be 
detrimental and our formalism does not address such a problem.  Nevertheless,
we find it encouraging that for the quite general model we have chosen,
where the data are required to simultaneously constrain thousands
of foreground parameters, the results look very good.

\section{Methodology}
We assume that the experiment measures the full sky in $\nu =
1,...,n_{\rm ch}$ channels, and model the (beam-deconvolved, 
spherical-harmonic transformed) map data as due
to the CMB, foregrounds and noise: 
\be 
\label{eqn:datamodel}
\Delta_{\nu lm} = \sum_i g_{i\nu} a_{ilm} + n_{\nu lm} 
\ee 
where $i$ runs over the components, ($i=0$ is CMB,
$i>0$ are the foregrounds) and $g_{i\nu}$ gives their frequency
dependence.  In the following we usually suppress all indices
and use a notation in which Eq.~\ref{eqn:datamodel} becomes: 
\be 
{\bf \Delta} = {\bf g}^\dagger {\bf a}+{\bf n}. 
\ee 
Throughout we assume
that the noise is spatially uniform, Gaussian-distributed, and
uncorrelated from channel to channel.  Therefore, 
${\bf W} \equiv 
<{\bf n} {\bf n}^\dagger>^{-1}$
is given by 
\be
W_{\nu l m,\nu\p l\p m\p} = w_\nu 
\delta_{\nu \nu\p} \delta_{l l\p} \delta_{m m\p}
\ee
where the weight-per-solid angle for map $\nu$, $w_\nu$, equals 
$B_{\nu,l}^2/(\sigma_\nu^2\Omega_{\rm pix})$, $\sigma_\nu$ is the standard error in the temperature
determination of a map pixel with solid angle $\Omega_{\rm pix}$, and
$B_{\nu,l}$ is the beam profile---which for a Gaussian beam with
full-width at half-max $\sqrt{8\ln 2}\sigma$ is given by
$exp(-(l\sigma)^2/2)$.  The beam-damping of the weight matrix is due
to the fact we are describing the noise in the {\it beam-deconvolved}
maps.

If we make specific assumptions about the statistics of the CMB and
foregrounds then we can determine how well we can measure the
parameters of those statistical distributions.  For simplicity and
specificity we assume the CMB and foregrounds to all be statistically
isotropic and Gaussian-distributed.  In this case a complete
statistical description of the data is given by the two-point
function:
\bea
\label{eqn:Mnoindices}
{\bf M}  & \equiv & \langle 
{\bf \Delta} {\bf \Delta}^\dagger \rangle \nonumber\\
 & = & {\bf g}^\dagger \langle {\bf a}{\bf a}^\dagger \rangle {\bf g}+ 
{\bf W}^{-1}.
\eea
If, in addition to statistical isotropy, we assume that each
of the foreground components are uncorrelated then we can write
\be
\langle a_{lm}^i a_{l\p m\p}^{i\p} \rangle = C_{il}\delta_{ll\p}
\delta_{mm\p}\delta_{ii\p}
\ee
and Eq.~\ref{eqn:Mnoindices} simplifies to (with indices restored):
\bea
\label{eqn:Mindices}
M^{\nu \nu\p}_{lm,l\p m\p} = \left[\sum_ig_{i\nu}g_{i\p \nu\p}C_{il}
+{1\over w_\nu}\delta_{\nu \nu\p}\right]\delta_{ll\p}\delta_{mm\p}.
\eea

Given the data, we could write down and calculate the posterior
probability distribution of the parameters, $C_{il}$, or any
other parameterization, $a_p$, of ${\bf M}$.  The posterior
is proportional to the product of the likelihood and the prior.
In the limit that the posterior distribution of $a_p$ is Gaussian,
the expectation value for the covariance matrix of the parameters is given
by the inverse of the ``posterior'' Fisher matrix, 
\bea
\label{eqn:fishmat}
F_{pp'} & \equiv &\langle {-\partial^2 \ln P_{posterior} \over \partial a_p
\partial a_{p'}} \rangle \nonumber\\
&=& {1 \over 2} {\rm Tr}\left({\bf M}^{-1}{\bf M}_{,p}{\bf M}^{-1}
{\bf M}_{,p\p}\right) +F^{\rm prior}_{p p\p}.
\eea
Note that the trace is a sum over $\ell$s, $m$s and $\nu$s.  ${\bf M}$
is block-diagonal with block size $n_{ch}$ by $n_{ch}$ so its inversion
is readily feasible.  The 
matrix $F$, or rather its inverse, is exactly what we want, the
expectation value of the covariance matrix of the 
parameters.  We are interested in calculating this parameter
covariance matrix for various parameter choices -- in particular
the $C_{il}$ -- as well as assumptions about their prior distributions.

We parameterize the (diagonal) prior as zero for $i=0$ and 
\be
F^{\rm prior}_{il,il} =(\alpha/C_{il})^2
\ee
for $i > 0$ where $C_{il}$ are the assumed actual power spectra, to be
discussed in the next section.  Note that if we take the foreground
$C_{il}$s to be a priori perfectly known ($\alpha \rightarrow \infty$),
then Eq. \ref{eqn:fishmat} gives the Fisher matrix for the 
Wiener filter method of foreground removal (TE96, BGP95), 
an explicit expression for which is in W98.  In the
absence of foregrounds it is equivalent to that in Bond, Efstathiou \&
Tegmark (1997, hereafter ``BET97'') and for a single channel
experiment it is equivalent to that given by Knox 1995 and by Jungman
\etal ~(1996).  Below we vary $\alpha$ to see quantitatively how the
strength of our prior assumptions determines the ability to measure
$C_{0l}$.

It is straightforward to generalize the above to include polarization
information.  Maps of the Q and U Stokes parameters can be decomposed
into two components, $a_{lm}^E$ and $a_{lm}^B$ (Kamionkowski \etal ~1997;
Zaldarriaga \& Seljak 1997), which are now in addition to the temperature
component $a_{lm}^T$.  In general, we can write the contribution from
each component as $a_{ilm}^b$ and the data in each channel as
$\Delta_{\nu,lm}^b$ where the superscript is either $T$, $E$ or $B$.
Then the covariance matrix for the data (Eq.~\ref{eqn:Mindices}) becomes
\bea
\label{eqn:Mpol}
M^{b \nu,b\p\nu\p}_{lm,l\p m\p} = 
\left[\sum_i g_{i\nu} g_{i\p\nu\p} C_{il}^{bb'}
+{1\over w^b_\nu}\delta_{bb\p} \delta_{\nu \nu\p}\right]\delta_{ll\p}\delta_{mm\p}
\eea
where $C_{il}^{bb'}$ equals 
$C_{il}^T \equiv \langle a^T_{ilm} {a^T_{ilm}}^* \rangle$ for $b=b'=T$,
$C_{il}^E \equiv \langle a^E_{ilm} {a^E_{ilm}}^* \rangle$ for $b=b'=E$,
$C_{il}^B \equiv \langle a^B_{ilm} {a^B_{ilm}}^* \rangle$ for $b=b'=B$, and
$C_{il}^C \equiv \langle a^E_{ilm} {a^T_{ilm}}^* \rangle$ for $b=T$, $b'=E$.
All other elements vanish.
Thus, while the matrix of Eq. ~\ref{eqn:Mindices} is 
block-diagonal in blocks of dimension $n_{ch}$, this matrix is 
block-diagonal in blocks of dimension $3n_{ch}$.  This approach
generalizese the multi-frequency Wiener filter error forecasting
of Bouchet \etal ~(1998, hereafter ``BPS''), who generalized 
the single-frequency, no foreground, treatment of Zaldarriaga \etal ~(1997).

We may also be interested in how well an individual mode can be
measured. 
The covariance matrix for the error in the minimum variance estimate 
of ${\bf a}$ is
\be
\label{eqn:fishmode}
\langle \delta {\bf a} \delta {\bf a}^\dagger \rangle = 
\left( {\bf g}^\dagger {\bf W} {\bf g} +{\bf W}^{\rm prior}\right)^{-1}
\ee
where we have assumed a prior probability for ${\bf a}$ that is
Gaussian-distributed with weight matrix ${\bf W}^{\rm prior}$.  For
example, we may wish to assume that foreground $i$ has variance
$C_{il}\delta_{ll\p} \delta_{mm\p}$ in which case $W^{\rm
prior}_{ilm,i\p l\p m\p} = 1/C_{il}\delta_{ll\p} \delta_{mm\p}$.  With
this prior, this is the variance given by the Wiener filter procedure.
Without the prior it is the variance given by the
pixel-by-pixel subtraction procedure of Dodelson
(1996) and also of Brandt \etal ~(1994) (except for their non-linear
parameter dependences).
When there are more foregrounds than channels,
${\bf g}^\dagger {\bf W} {\bf g}$ is singular and therefore addition
of a prior is necessary to make $\langle \delta {\bf a}\delta {\bf
a}^\dagger \rangle$ finite.  For more flexibility in the prior
choice later, we define $\beta$ so that $W^{\rm
prior}_{ilm,ilm} = \beta/C_{il}$.  Note that Eq.~\ref{eqn:fishmode} does not
assume anything about the statistical properties of the foregrounds
and CMB---except through the prior, which we have explicitly assumed
to be Gaussian.

\section{Foreground Models}

Our foreground model is based on that developed for the Planck Phase A
proposal (Bersanelli \etal ~1996) and updated in the HFI and LFI
instrument proposals.  We refer the interested reader to these
proposals and to Bouchet and Gispert (1998,
hereafter ``BG98'').  Below we briefly describe our model, with an
emphasis on the modifications and additions we have made.  In all
cases, these alterations make the model more pessimistic.

\subsection{Galactic}

Analyses of the DIRBE (Diffuse Infrared Background Explorer) and 
IRAS (Infrared Astronomy Satellite) Sky Survey Atlas maps 
have determined the shape of
the dust power spectrum to be $C_l \propto l^{-3}$ (Gauttier \etal
~1992; Wright 1997) or $C_l \propto l^{-2.5}$ (Schlegel \etal ~1998).
We assume $C_l \propto l^{-2.5}$ since it is the
more pessimistic choice, given that we normalize at large angles.

We take the same $C_l$ shape for the free-free power spectrum because
both the dust intensity and free-free are expected to be from the same
warm interstellar medium.  Indeed, there is strong observational
evidence for a correlation (Kogut \etal 1996, Leitch \etal 1997, de
Oliveira-Costa \etal 1997, Jaffe \etal 1998).  Note, however, that we assume no
cross-correlation between free-free and dust, because any correlation
makes the foreground separation easier.  The same shape is also taken
for synchrotron radiation.

We choose amplitudes and frequency dependences for the galactic
foregrounds consistent with the Kogut \etal ~(1996) analysis of
DMR, DIRBE and Haslam maps.  We take the antenna temperatures
of the free-free and synchrotron to vary with power-law indices
-2.9 and -2.16, respectively.  For the dust we assume a $\nu^2$
emissivity dependence and a single component with $T=18K$.  

Draine and Lazarian (1997) have proposed an alternative explanation
to the observed correlation between dust and 30~GHz to 90~GHz radiation.
They propose that the rotational emission from spinning dust
grains, greatly increases the emission in the 10~GHz to 100~GHz range
above what one expects from the vibrational emission.  
We have not included this component of dust emission in our model.
Instead, we include something worse -- a component with spectral
shape similar to the ``anomalous'' emission, but which has no correlations
with the dust.  Again, this is more pessimistic than the strong
correlation expected in a realistic model.

\begin{figure}[bthp]
\plotone{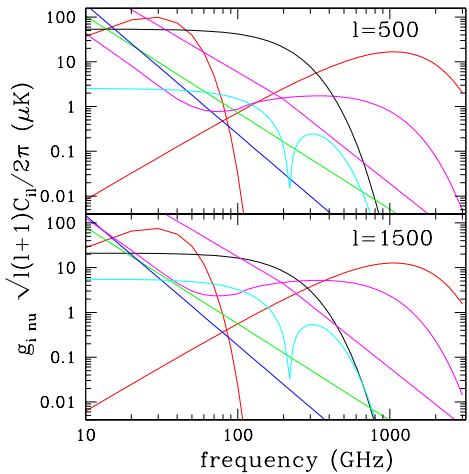}
\caption[foreground model]{\baselineskip=10pt 
The frequency-dependent rms antenna temperature,  
$g_{i\nu}\sqrt{l(l+1)C_{il}/(2\pi)}$
evaluated at $l=500$ (top panel) and $l=1500$ (lower panel) 
for standard cdm CMB (black), dipole and thermal emission dust (both red), 
free-free (green), 
synchrotron (blue), SZ (cyan), and radio and FIR point sources (both magenta).  
}
\label{fig:model}
\end{figure}

\subsection{Extragalactic}
Extragalactic contributions to the microwave sky include inverse
Compton scattering of CMB photons by hot gas in clusters (the thermal 
Sunyaev-Zeldovich (SZ) effect),  the Far Infrared Background (FIRB)
and radio point sources.  

Following Tegmark and Efstathiou, we model the contribution from a
diffuse background of unresolved radio point sources as having an
intensity, $I(\nu) \propto \nu^{-\alpha}$ with a white noise angular
power spectrum ($C_\ell$ independent of $\ell$).  Deviations from
white noise due to clustering are observed to be neglible at 1.5GHz
(TE96; Tofollati \etal ~1998, herafter ``To98'').  Below 200 GHz, we
take $\alpha=0$ but above 200 GHz we introduce a break to $\alpha=0.7$
as suggested by To98.  We adopt this break in our spirit of
pessimism because, despite decreasing the brightness of this
contaminant, it actually makes determination of the CMB more
difficult.  This is due to the fact that with the break, the spectral
shape more closely resembles that of the CMB.  

We are actually considering the power spectrum of the sources which
remain after some cut is done of clearly contaminated pixels, e.g.,
those above a $5\sigma$ threshhold where $\sigma^2$ is the variance in
the map.  Thus the amplitude depends both on the number-count
distribution and on the level of flux cut that is used.  Although this
flux cut will vary for maps at different frequencies and from
different instruments, we choose to fix it at 1 Jy.  We view this as
quite conservative since the typical level for all the Planck maps is
about $\sigma$ = 0.1 Jy.  This is according to Tegmark \& de Oliveira-Costa
(1998) who used the point-source model of To98 and included the effect
of reduction in $\sigma$ that one can achieve by applying a Wiener
filter.  The values of $\sigma$ for the MAP maps should not differ by
more than a factor of 2 from those for the LFI.

For the amplitude of the FIRB we
rely on the estimates of BG98 which are derived from
the model of FIR point sources of Guiderdoni \etal ~(1998).  
This model has succesfully predicted source counts in a
wide wavelength range, from 15 to 850 microns (see BG98
and references there in).
The mean frequency dependence of the model is shown in
Fig.~\ref{fig:model}.  Bouchet and Gispert (1998) have shown that this
frequency dependence has only slight spatial variations, lending
credence to our modeling of it as a frequency dependence times a fixed
brightness spatial template.  We assume clustering is unimportant and
therefore the spatial power spectrum has the same shape as we
have assumed for radio point sources:  $C_l$ is a constant.

CMB photons moving through a hot gas of electrons have their frequency
shifted as they Compton scatter, leading to the generation of
anisotropy with a non-thermal spectrum.  This Sunyaev-Zeldovich (SZ) effect
can also be treated as an additive foreground, with the
frequency-dependence of a Compton $y$ distortion.  Calculations of the
power spectrum of this foreground component, assuming a
Press-Schechter distribution of clusters with masses greater than some
cutoff have been done (Aghanim \etal ~1996, herafter A96;Atrio-Barandella 
\& Mucket 1998).  We use
the results of A96 for the $\Omega=1$ cosmology.  Their power
spectrum is well-fit in the range of interest by $C_l = a(1+l_c/l)$
where $l_c=1190$ and $a$ is such that
$l(l+1)C_l/(2\pi) = 5.3 \muk^2$ at $l=1500$ in the Rayleigh-Jeans
(low frequency) limit (see Fig.~\ref{fig:model}).  Modelling of this
contribution will soon be improved by replacement of the use of
Press-Schechter with N-body/hydro simulations.

\subsection{Spectral Shape Uncertainty}

Implicit in our formalism is that the frequency dependence of the
foregrounds is known perfectly and has no spatial variation.  However,
we can allow for some degree of spatial dependence of the spectrum as
follows.  Consider foreground $i$ with mean power-law frequency
dependence, $\beta$, and deviation $\delta \beta_{lm}$.  Then, the
signal contribution to the data, $\Delta_{\nu lm}$ from component $i$
is
\be 
a_{ilm}(\nu/\nu_0)^{\beta+\delta \beta_{lm}} \simeq
a_{ilm}(\nu/\nu_0)^{\beta} + a_{ilm}\delta \beta_{lm} (\nu/\nu_0)^{\beta} 
\ln(\nu/\nu_0).
\ee
Thus we can treat radiation from a component with spatially 
varying spectral index as due to two components with amplitudes
$a_{ilm}$ and $a_{ilm} \delta \beta_{lm}$, which will, in general,
be correlated.  For simplicity we have modeled these additional
components as uncorrelated with the parent component and taken
$\langle a_{ilm} \delta \beta_{lm} a_{il\p m\p}\beta_{l\p m\p}
\rangle= C_{il} \langle \delta \beta^2 \rangle$.  We have assumed
$\langle \delta \beta^2\rangle = 0.25$ for the rotating small dust
grains, dust, and synchrotron with the same prior as
used on other foregrounds.  TE96 also considered using extra
components to model spatial dependence of the spectral shape.  For
an alternative approach, see Tegmark (1997).

\subsection{Foreground Polarization}

Precious little is known about the polarization of foregrounds.
For a review, see Keating \etal ~(1998).
Extrapolation from observations at low-frequency ($\la 1$ GHz) are
complicated by Faraday rotation along the line-of-sight, which is
negligible at higher frequencies.  Measurements at higher frequencies
are in the galactic plane in dense star-forming regions (Hildebrand \&
Dragovan 1995) and are not expected to be representative of the
statistics at high latitude.  We make the same assumptions about foreground
polarization as BPS.   They neglect polarization in all foregrounds
except for synchrotron and dust.  For the synchrotron, they
take $C_l^E = 0.2C_l^T$ and for the dust they take the
model of Prunet \etal ~(1998, hereafter ``PSB'') 
(see also Sethi \etal ~(1998)). 
It must be kept in mind that the PSB calculation
relies on indirect arguments and is therefore quite uncertain, as is
the synchrotron model, as the authors readily admit.  

\section{Application to Planned Experiments}

\subsection{Temperature}
\begin{figure}[bthp]
\plotone{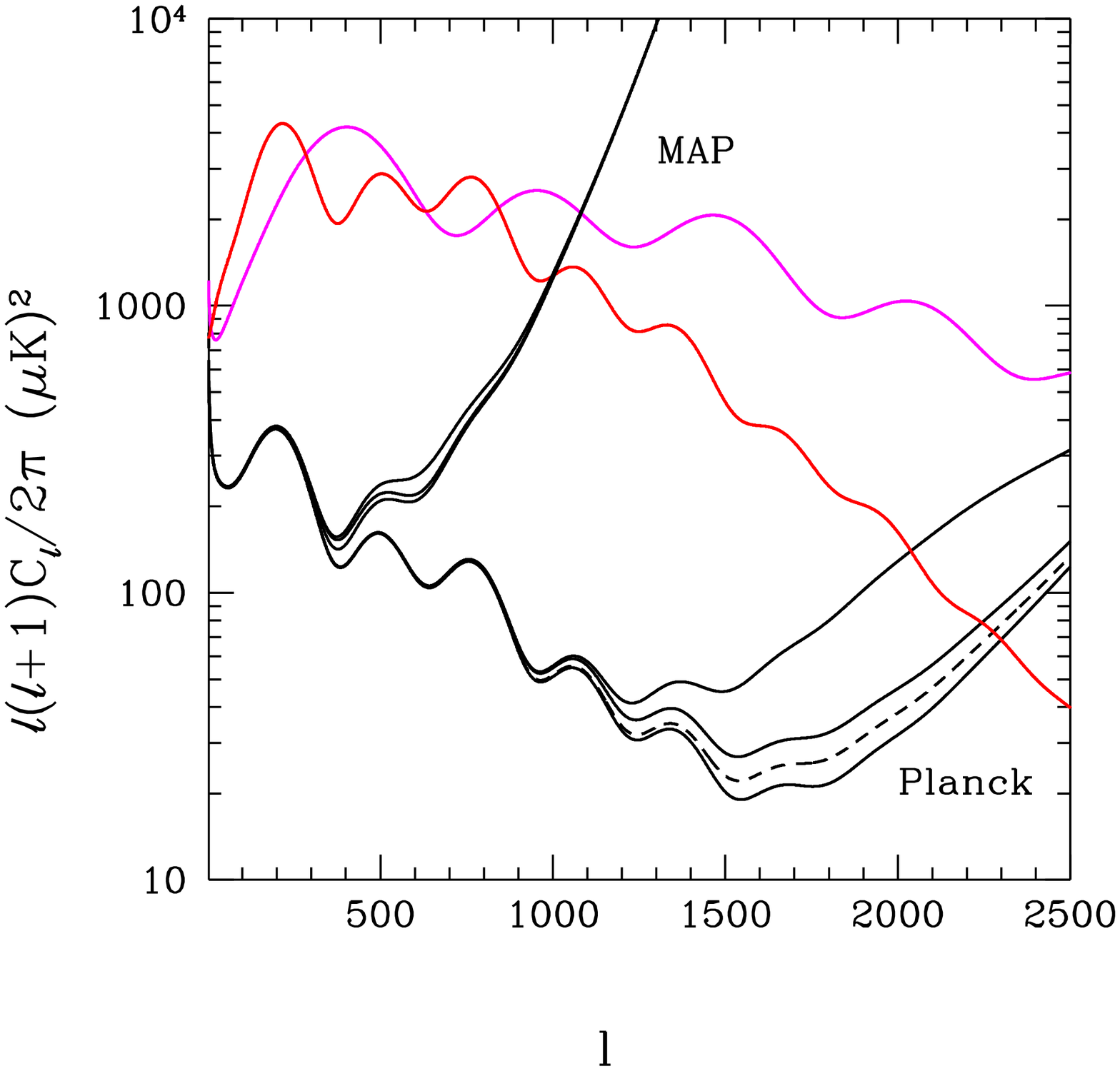}
\caption[powspec errors]{\baselineskip=10pt 
MAP uncertainties for $\alpha = 0.1$ on all foreground components (top curve),
$\alpha=0.1$ on all foreground components except for 
radio point sources and SZ, for which $\alpha=1.1$ (second to top
curve).  The lowest
uncertainty curve is identical for $\alpha=\infty$ and no 
foregrounds.  Planck uncertainties for 
$\alpha = 0.1$ on all foreground components except radio
point sources and the FIRB for which $\alpha=1.1$
(highest curve); $\alpha=\infty$ (middle solid curve)
the no foreground case (bottom curve), and   
same as the top curve but with the FIRB and radio point source 
power spectra reduced by sixteen (dashed curve).  
With the FIRB and radio point source power spectra reduced by a factor
of 16, the $\alpha=\infty$ case is identical with the no foreground
case.   
}
\label{fig:dclt}
\end{figure}

In Fig.~\ref{fig:dclt} one can see that MAP's ability to measure
the power spectrum is not significantly affected
by the foregrounds below $\ell \simeq 500$.  Going to smaller
values of $\ell$ we have greater frequency coverage, and greater
ratio of signal to instrument noise. 
The only thing that gets slightly
worse as $\ell$ decreases is the relative amplitude of the
galactic foreground power spectra, but this effect is overwhelmed
by the others.  Of course going to higher $\ell$ we have
less frequency coverage and a smaller ratio of signal to
instrument noise.  The galactic
foregrounds still do not become a problem though since their
relative power continues to decrease.  

What does become a concern at higher $\ell$
are foregrounds with rising angular power spectra:  radio point sources
and the thermal Sunyaev-Zeldovich effect from galaxy clusters.
These alone are responsible for the deviation of $\Delta C_l$ from the
no foreground case, visible in Fig. \ref{fig:dclt}.  

The impact of the Sunyaev-Zeldovich component is worth exploring more.
It is quite possible that the actual amplitude is ten times larger
than in our baseline model. The A96 calculation ignores
the contribution from filaments---which may actually dominate the
contribution from the clusters, and it ignores the clustering of the
clusters.  If we increase the power by a factor of 10, and relax the
prior on it to $\alpha = 0.1$ from $\alpha=1.1$, $\Delta C_l$ doubles
in the range from $l=400$ to $l=700$.  On the other hand, if we
increase the power by a factor of 10, and do not relax the prior,
$\Delta C_l$ only increases by a few per cent.  What we learn from
this is that having some constraints on the power spectrum of the SZ
component can be just as important as the actual amplitude.

The usefulness of prior knowledge of the SZ $C_l$ is encouraging.
It suggests that the analysis of MAP data can 
profit significantly from accurate theoretical
predictions of the statiscal properties of the SZ component.  
It also suggests that measurements of the SZ component in much
smaller regions of the sky, which roughly constrain the power
spectrum, can be beneficial to the analysis of the full-sky MAP data.
Such analyses should be possible from combining MAP data with
datasets from higher frequency instruments such as TopHat\footnote{
TopHat home page:
{\tt http://\linebreak[1]topweb.gsfc.nasa.gov}}
and
BOOMERANG\footnote{
BOOMERANG home page:\\ {\tt http://\linebreak[1]astro.caltech.edu/\~{}mc/boom/boom.html}}, which by themselves will be extremely interesting CMB datasets.

Planck's ability to measure the power spectrum is not significantly
affected by the foregrounds below $\ell \simeq 1200$.  At higher
$\ell$, the frequency coverage reduces, the noise in each channel
increases and the SZ, radio and FIRB components increase in amplitude.
Unlike for MAP, SZ is not important because in the HFI frequency
range, SZ is easily distinguished from CMB; there is even the null
at 217 GHz.  However, the radio point sources and FIRB are a
concern.  There is strong dependence on the prior.  Even with moderate
prior information ($\alpha=1.1$ on these two components), $\Delta
C_{0l}$ is 3 times larger than the no foreground case.  With an
infinite prior this reduces to a much less significant factor of about
1.2.  The situation is greatly improved if the flux from the two 
backgrounds of unresolved sources is a factor of 4 less in
amplitude (16 in $C_l$) than we have assumed.  This is not unlikely
since our assumed flux cut of 1 Jy is about 20 times the level of
confusing noise, calculated by Tegmark \& de Oliveira-Costa (1998), 
in the (post-Wiener filtering)
143 GHz, 217 GHz and 353 GHz HFI maps, and is therefore an extremely
conservative $20\sigma$ cut.  Thus, we also show the results with our
input power spectrum for point sources, and the FIRB each reduced by a
factor of 16 as the dashed line in Fig.~\ref{fig:dclt}.

We see that with only the use of a moderate amount of prior
information, the errors on the $C_{0l}$s here are not qualitatively
different from the no-foreground results.  The conclusions of those
forecasting cosmological parameter errors would not be qualitatively changed
by including the effect of the foregrounds as modelled here.

\begin{figure}[bthp]
\plotone{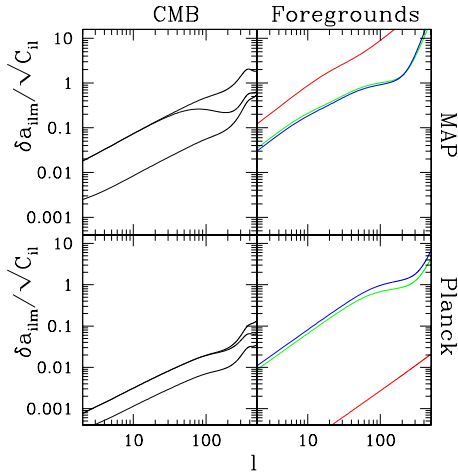}
\caption[fourpanel]{\baselineskip=10pt 
MAP (top) and Planck (bottom) component map uncertainties
expressed as 
$\delta a_{ilm}/\sqrt{C_{il}}$ for the CMB (left) and the 
three galactic foregrounds (right).  No other
foregrounds were included in the calculation of the uncertainty.
On the left, the three cases are pixel by pixel subtraction 
($\beta=0$, top curve), Wiener filtering ($\beta=1$, middle curve) and 
no foregrounds (bottom curve).  On the right the three cases
are dust (red), free-free (green) and synchrotron (blue) with
a $\beta=1$ prior
applied to all the components except for the CMB and the component in
question.  
}
\label{fig:4panel}
\end{figure}

If galactic foregrounds are well-described by the model used
here, then they will not have significant impact on the
primary science goals of MAP and Planck.  That is perhaps
the most robust conclusion to draw from the above.  This
is not to say that these foregrounds do not have their impact on how well
the CMB can be measured.  The left side of Fig.~\ref{fig:4panel}
shows how the foregrounds affect the uncertainties 
in $a_{lm}^{\rm CMB}$.  As long as $\delta a_{lm}^{\rm CMB}/\sqrt{C_l}
< 1$ then sample-variance dominates the errors in $C_\l$.  As 
can be seen in the figure, this inequality holds out to at least
$l = 500$, except for MAP in the case of pixel-by-pixel subtraction
($\beta = 0$, or no use of prior information).

\subsection{Polarization}
The CMB is expected to be polarized at a level of about 10\% of the
anisotropy.  The polarization foregrounds are no where near
as well-understood and explored as the temperature foregrounds.  
However, taking some initial guesses at the polarization foregrounds
we find the outlook for CMB polarization measurement by MAP and
Planck to be fairly bright.  The reason being that, once again,
there is a window in frequency space where the CMB is the dominant
contributor to spatial variations in polarization.  

This window does not necessarily exist across the
entire polarization power spectrum, and in particular may disappear at
low $l$. This is unfortunate since the two most interesting features
in the polarization power spectra are the bump at $l \simeq
2\sqrt{z_{ri}}$ where $z_{ri}$ is the redshift of reionization and the
B-mode power spectrum due to tensor and vector modes (Kamionkowski \etal 1997,
Seljak \& Zaldarriaga 1997) which also peaks at low $l$.

Here we focus on the reionization bump.  To study sensitivity to it
we have not implemented Eq.~\ref{eqn:Mpol}.  Instead we have
ignored cross-correlations between temperature and polarization
so that Eq.~\ref{eqn:Mindices} is applicable with appropriate
substitutions (e.g., $C_{il} \rightarrow C_{il}^E$).  In general
the cross-correlations improve the constraints on the polarization
power spectrum (BPS) but that shouldn't be the case here since 
the reionization bump is a feature solely of the polarization maps
and does not show up in cross-correlation with temperature maps.

For standard CDM with an optical depth to Thomson scattering of
$\tau=0.1$, Planck measures the reionization feature with cosmic
variance precision (although HFI alone does not and neither does MAP).
At larger $\ell$, where the signal is large, the foregrounds, as
modelled here, have no significant impact on the ability of either of
the satellites to measure the CMB polarization power spectrum.  Our
infinite foreground prior (or Wiener filter) results are in agreement
with the Wiener filter results of BPS.

\begin{figure}[bthp]
\plotone{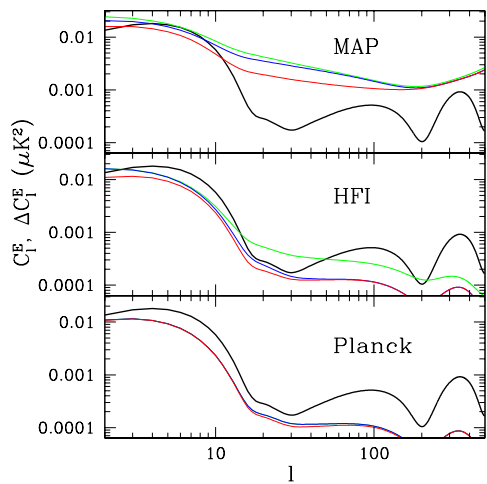}
\caption[powspec errors]{\baselineskip=10pt 
$C_\ell^E$ vs. $\ell$ for standard cdm (thick black line) with optical depth
to Thomson scattering of $\tau=0.1$ and expected uncertainties under different
assumptions.  The assumptions are, from top to bottom in each panel,
no foreground prior (green), infinite foreground prior (blue) and 
no foregrounds (red).
}
\label{fig:dclp}
\end{figure}

\section{Discussion}

We have presented a method to calculate the sensitivity to
the CMB and its power spectrum given multi-resolution, 
multi-wavelength observations of a sky that consists of
multiple foreground contributions\footnote{IDL programs implementing
this procedure are available from the author.}.  The 
applications to MAP and Planck have allowed for much greater freedom
in the behavior of the foregrounds than did previous analyses
(TE96, BG98).  Despite this extra freedom, the conclusions
are similar---that foregrounds are not likely to qualitatively
affect the uncertainties that can be achieved on cosmological
parameters.  Similar conclusions have been reached by 
de Oliveira-Costa \etal (1998).  

Our approach has not fully taken into account the non-Gaussianity of
the foregrounds, spatial dependence of the spectrum of each component,
uncertainty in the spectral shapes, and unknown components (e.g., a
population of points sources whose spectra peak at 90 GHz).  For these
reasons it is difficult to conclude with certainty that the
foregrounds will not qualitatively affect the determination of
cosmological parameters.  However, a very important reason for our
rosy conclusions is a very simple fact: for most of the multipole
moments measured by a given experiment, the quality of the CMB map can
be highly degraded, without having any impact on the quality of the
power spectrum.  Thus, any effect we have not included here has to
overcome this hurdle in order to be important.

Non-gaussianity is both a friend and a foe.  We have already exploited
it here in assuming that the brightest points sources could be
identified with threshhold cuts and removed.  However, it can
present a challenge to the above sample-variance argument  
if it resulted in the errors in each $a_{lm}$ being
strongly correlated with the errors in $a_{lm\p}$, in such a way
that they did not beat down with many averagings.  One can think
of this as an effective reduction in the number of independent
modes in the foreground (Eisenstein 1998).    
However, we expect that small-scale behavior in patches
of the sky sufficiently separated to be decorrelated.  Hence we
do not expect the mode number reduction to be large,
though further investigation of effects of non-Gaussianity is
clearly warranted.

We have also negected things that will improve estimation of the
CMB from MAP and Planck data, such as the use of maps at other
frequencies, e.g., DIRBE, IRAS and FIRST (which will fly with Planck).
Assumptions about the smoothness of foreground power spectra are
also reasonable and could significantly reduce our error forecasts
at high $l$, by extending the information gained at lower $l$
where there is greater frequency coverage.  

It is clear though, that even if foregrounds do not do anything more
than double the errors on cosmologcial parameters, the determination
of the exact size of the error bars will probably be dominated by
foreground considerations.  Small patches of the sky will be analyzed
separately, with those appearing the cleanest given more weight.
Foreground model residuals will be agressively sought.  Thus the study
of foregrounds remains very important.  We close by listing the
following improvements in our understanding of foregounds which could
prove to be extremely beneficial:

\bul More accurate theoretical calculation of the statistics of the
SZ component.  Our positive conclusions for MAP depend on the
amplitude of the SZ power spectrum
and on how well that power spectrum
can be determined {\it a priori}.  We have shown that having
a prediction of $C_l^{\rm SZ}$ good to about a factor of 2
(which would justify our use of $\alpha=1.1$) is enough to
keep $\Delta C_l$ within about ten per cent of the no foreground case,
even if $C_l^{\rm SZ}$ is ten times larger than the A96 calculation.

\bul Higher frequency complements to MAP, such as are coming from
balloon flights (e.g., TopHat and \\ 
BOOMERANG).  Even coverage
of just a few per cent of the sky, can be used to characterize
the level of contamination in the rest of the sky.

\bul A point source survey near 90 GHz (see Gawiser \etal ~1998).  

\bul Further development of methods for removing non-Gaussian
foregrounds and understanding of resulting CMB uncertainties.

\bul A full-sky, high resolution H$\alpha$ survey, since this
is a tracer of free-free.  Useful steps in this direction
are already being made with a survey in the North with one degree
resolution (Reynolds \etal 1998) and one in the South with 0.1 degree 
resolution (McCullough \etal 1998) nearing completion.

\bul Measurements of high galactic latitude dust and synchrotron polarization. 

\acknowledgments
I thank K. Ganga, A. Jaffe and J. Ruhl for useful conversations,
as well as the organizers and participants of the Sloan Summit on Microwave
Foregrounds, who have informed the above discussion.  I acknowledge
the use of CMBFAST (Seljak \& Zaldarriaga 1996).


\end{document}